\documentclass[dvips]{article}

\usepackage{icrc2011}

\title{AstroParticle Physics at the Highest Energies}

\newcommand{\etal}{\MakeLowercase{\textit{et al. }}} 
\shorttitle{author \etal paper short title}

\def\lsim{\mathrel{\rlap{\lower4pt\hbox{\hskip1pt$\sim$}}
    \raise1pt\hbox{$<$}}}                
\def\gsim{\mathrel{\rlap{\lower4pt\hbox{\hskip1pt$\sim$}}
    \raise1pt\hbox{$>$}}}                

\authors{Angela V. Olinto}
\afiliations{Department of Astronomy and Astrophysics, 
Kavli Institute for Cosmological Physics, \\
The University of Chicago, Chicago, IL 60637, USA }
\email{olinto@kicp.uchicago.edu}

\abstract{Recent international efforts have brought us closer to unveiling the century old mystery of the origin of cosmic rays. Cosmic ray, gamma ray, and neutrino observatories are reaching the necessary sensitivity to study the highest energy cosmic accelerators and to begin the use of cosmic particles to study particle interactions above laboratory energies. The number of known gamma-ray sources has increased by orders of magnitude. Possible cosmic ray sources  have narrowed down with the confirmation of an ankle and the GZK-like spectral feature at the highest energies. Anisotropies in the distribution of arrival directions of cosmic rays at intermediate energies show a complex local neighborhood of the Galaxy. At the highest energies the dawn of particle astronomy is still challenging while composition related measurements point to a change in the composition or the interaction of cosmic rays at ultrahigh energies. A clear resolution of the ultrahigh energy mystery calls for a significant increase in statistics of cosmic ray and neutrino observations. 
}

\begin{document}
\maketitle

\section{Introduction}

\begin{figure}[!t]
\centerline{\includegraphics[angle=270,width=0.55\textwidth]{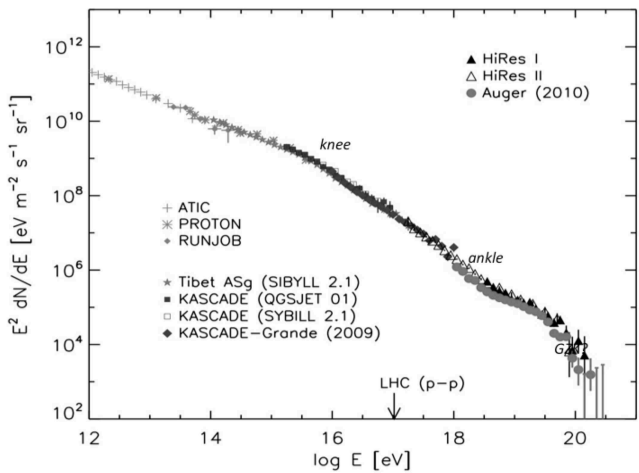}}
\caption{All particle cosmic ray flux multiplied by $E^2$ observed by ATIC \cite{Ahn08}, Proton \cite{Grigorov71}, RUNJOB \cite{Apanasenko01}, Tibet AS-$\gamma$ \cite{Chen08}, KASCADE \cite{Kampert04}, KASCADE-Grande \cite{KASCADE-Grande11}, HiRes-I \cite{Abbasi09}, HiRes-II \cite{Abbasi08}, and Auger \cite{Abraham10} (adapted from \cite{KKAO11}).  LHC energy reach of $p-p$ collisions (in the frame of a proton) is indicated for comparison.}
\label{CRspec}
\end{figure}

The mystery of the origin of cosmic rays is celebrating its 100th, anniversary in 2012. 
As shown in figure \ref{CRspec}, the study of this striking non-thermal spectrum requires a large number of instruments to cover over 8 orders of magnitude in energy and 24 in flux. Galactic accelerators are likely responsible for the dominant component of cosmic rays observed on Earth, given the containment of lower energy cosmic rays by the Galactic magnetic field. Recent increase in gamma-ray observations (see, e.g., \cite{Aharonian}) have opened the possibility that the origin of these Galactic cosmic rays will be soon identified. Gamma-ray observations from GeVs to 100s of TeV  show at least 10 populations of gamma-ray generating astrophysical accelerators in the Universe. The main challenge now is to identify the hadronic accelerators among this list of sources where the leading candidate continues to be shock acceleration in supernova remnants. Gamma-ray and neutrino telescopes together with intermediate energy cosmic ray observatories are likely to determine the origin of Galactic comic rays  in the near future, but the origin of the extragalactic component at the highest energies is still quite puzzling. 

Models for these unknown extragalactic cosmic ray accelerators are challenged by the extreme energies of these particles, observed to reach 100s of EeV (1 EeV $\equiv 10^{18}$ eV), while their observation is difficult due to the very low flux of these extreme events, below 1 particle per km$^2$ per century. 

Before becoming a mainly extragalactic cosmic ray population at the highest energies, a transition from Galactic to extragalactic cosmic rays should occur somewhere between the {\it knee} of the cosmic ray spectrum at a few PeV ($\equiv 10^{15}$ eV) and the {\it ankle} at a few EeV. These features are shown in figure \ref{CRspec}. The spectral shape and composition of this transition will help illuminate the possible sources of both Galactic and extragalactic cosmic rays as discussed in section \ref{section:transition}.

Above a few EeV,  the so-called ultrahigh energy cosmic rays (UHECRs) are most likely extragalactic (recent reviews can be found in \cite{KKAO11,Letessier11}). These are observed to reach energies that exceed $10^{20}$ eV posing some interesting  questions: Where do they come from?  What kind of particles are they? What is the spatial distribution of their sources? How are they accelerated to such high energies? What do they tell us about these extreme cosmic accelerators? How strong are the magnetic fields that they traverse on their way to Earth? How do they interact with the cosmic background radiation? What secondary particles are produced from these interactions? What can we learn about particle interactions at these otherwise inaccessible energies? 

Below, we summarize recent observations of the spectrum, composition, and the search for anisotropies in the sky distribution of UHECRs (section \ref{section:observation}). The spectral  shape supports the notion that sources of UHECRs are extragalactic including a spectral feature  of a steep decline in flux above about 30 EeV (see  figure \ref{uhecrSpec}). This feature can be explained by the maximum energy of cosmic rays accelerators, $E_{\rm max}$,  or can be due to the effect of interactions between extragalactic cosmic rays and the cosmic background radiation, named the Greisen-Zatsepin-Kuzmin (GZK) effect \cite{G66,ZK66}. Hints of anisotropies have been reported by the leading UHECR observatory, the Pierre Auger Observatory, while composition indicators from shower development observations argue for a transition to a heavier component from a few EeV up to 40 EeV \cite{Abraham:2010yv}. Heavy nuclei dominated injection models are quite rare in the astrophysical literature of candidate sources (see section \ref{section:sources}) and if iron is the main component at the highest energies, Galactic magnetic fields should wash out most anisotropic patterns around 60 EeV. Another possible interpretation of the observed shower development properties is a change in hadronic interactions above 100 TeV center of mass (TeV $\equiv 10^{12}$ eV), an order of magnitude higher energy than reached by the Large Hadron Collider (LHC) at CERN. 

A new puzzle is born: an injection at the source dominated by heavy nuclei is astrophysically unexpected, while significant changes in hadronic interactions represent novel particle physics  above 100 TeV center of mass. With a significant increase in the integrated exposure to cosmic rays above 60 EeV, next generation UHECR observatories may reach the sensitivity necessary to achieve charged particle astronomy. In addition, UHE neutrino and photon observations  can further illuminate the workings of the Universe at the most extreme energies.

\section{UHECR observations}\label{section:observation}

After many decades of efforts to discover the origin of cosmic rays, current observatories are now reaching the necessary exposure to begin unveiling this longstanding mystery (see figure~\ref{ExposFutur} for a history of the exposures of the largest observatories). The first detection of UHECRs dates back to \cite{Linsley63}, but it was only during the 1990s that  international efforts began to address these questions with the necessary large-scale observatories. The largest detectors operating during the 1990s were the Akeno Giant Air Shower Array (AGASA), a 100 km$^2$ ground array of scintillators in Japan \cite{Chiba92}, and the High Resolution Fly's Eye (HiRes)  a pair of fluorescence telescopes that operated in Utah until 2006 \cite{Boyer02}. During their lifetimes, AGASA reached an exposure of $1.6 \times 10^3$ km$^2$ sr yr while HiRes reached twice that. To date, the highest energy recorded event was a 320 EeV fluorescence detection \cite{Bird:1994} by the pioneer fluorescence experiment Fly's Eye \cite{Baltrusaitis85}.

Completed in 2008, the Pierre Auger Observatory is the largest observatory at present \cite{Abraham04}. Constructed in the province of Mendoza, Argentina, by a collaboration of 18 countries, it consists of a 3,000 km$^2$ array of water Cherenkov stations with 1.5 km spacing in a triangular grid overlooked by four fluorescence telescopes. The combination of the two techniques into a hybrid observatory maximizes the precision in the reconstruction of air showers, allowing for large statistics with good control of systematics. The largest observatory in the northern hemisphere, the Telescope Array (TA), is also hybrid \cite{TA}. Situated in Utah, it covers 762 km$^2$ with scintillators spaced every 1.2 km overlooked by three fluorescence telescopes.

\subsection{Spectrum}

\begin{figure}[!t]
\centerline{\includegraphics[angle=270,width=0.55\textwidth]{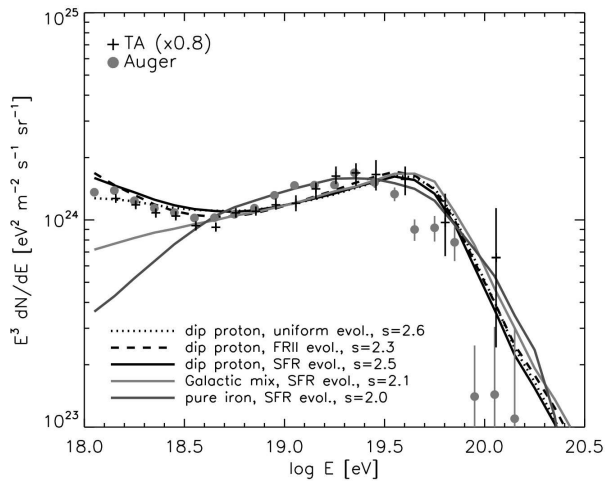}}
\caption{Flux of UHECRs multiplied by $E^3$ versus energy from the Auger Observatory \cite{AugerSpecICRC} and  the Telescope Array \cite{TA} (the TA absolute energy has been multiplied by 0.8). The displayed error bars are statistical errors while the reported systematic error on the absolute energy scale is 22\%. Overlaid are simulated spectra obtained for different models of the Galactic to extragalactic transition (ankle or ``dip'' transition) and different injected chemical compositions (pure protons, Galactic mix, and pure iron) and spectral indices, $s$ from 2 to 2.6 (adapted from \cite{KKAO11}). }
\label{uhecrSpec}
\end{figure}

The observed cosmic ray spectrum (figure \ref{CRspec}) can be described by a broken power law, $E^{-s}$, with spectral  index $s = 2.7$ below the {\it knee} at $\sim$ 1 PeV ($=10^{15}$ eV) and $s\simeq 3$ between the knee and the ankle around 3 EeV. Above the ankle, recent observations  reveal a spectrum whose shape supports the long-held notion that sources of UHECRs are extragalactic. As shown in Figure \ref{uhecrSpec}, the crucial spectral feature recently established at the highest energies is a steeper decline in flux above about 30 EeV. This feature was first established by the HiRes Observatory \cite{Abbasi09} and confirmed with higher statistics by the Pierre Auger Observatory \cite{Abraham:2008ru}. This steep decline in flux is reminiscent of the effect of interactions between extragalactic cosmic rays and the cosmic background radiation,  the GZK effect, which causes cosmic ray protons above many tens of EeV to lose energy via pion photoproduction off cosmic backgrounds while cosmic ray nuclei photodissociate. This feature was not seen in earlier observations with the AGASA array \cite{Takeda98}. Data from the Auger Observatory \cite{AugerSpecICRC} and preliminary data from the Telescope Array \cite{TA} are shown in the  figure  \ref{uhecrSpec}. The observations agree well given an overall energy re-scaling of 20\% which is within the systematic errors in the absolute energy scale of 22\%.

Another important feature shown in figure \ref{uhecrSpec}  is the hardening of the spectrum at a few EeV, called the {\it ankle}, which may be caused by the transition from Galactic to extragalactic cosmic rays or by propagation losses if UHECRs are mostly protons.

Figure \ref{uhecrSpec} shows the observed spectrum fit by different models of UHECR sources (adapted from \cite{KKAO11}). In the mixed composition and iron dominated models   \cite{Allard07}, the ankle indicates a transition from Galactic to extragalactic cosmic rays, the source evolution is similar to the star formation rate (SFR),  and the injection spectra are relatively hard (power law index $s\sim 2 -2.1$). In the proton dominated models in the figure, the ankle is due to pair production propagation losses \cite{BG88}, named ``dip transition models" \cite{BGG06}, and the injection spectra are softer for a wide range of evolution models. Models with proton primaries can also fit the spectrum with harder injection with a transition from Galactic to extragalactic at the ankle.

The confirmed presence of a spectral feature similar to the predicted GZK cutoff, settles the question of whether acceleration in extragalactic  sources can explain the high-energy spectrum, ending the need for exotic alternatives designed to avoid the GZK feature. However, the possibility that the observed softening of the spectrum is mainly due to the maximum energy of acceleration at the source, $E_{\rm max}$,  is not as easily dismissed. A confirmation that the observed softening {\it is} the GZK feature,  awaits supporting evidence from the spectral shape (at energies above 100 EeV), anisotropies (which are expected above GZK energies), composition, and the observation of produced secondaries such as neutrinos and photons.

\subsection{Anisotropies}

The landmark measurement of a flux suppression at the highest energies encourages the search for sources in the nearby extragalactic Universe using the arrival directions of trans-GZK cosmic rays (with energy above $\sim$ 60 EeV). Above GZK energies, observable sources must lie within about 100 Mpc, the so-called GZK horizon or GZK sphere. At  trans-GZK energies, light composite nuclei are promptly dissociated by cosmic background photons, while protons and iron nuclei  may reach us from sources at distances up to about 100 Mpc. Since matter is known to be distributed inhomogeneously within this distance scale, the cosmic ray arrival directions should exhibit an anisotropic distribution above the GZK energy threshold, provided intervening magnetic fields are not too strong. At the highest energies, the isotropic diffuse flux from sources far beyond this GZK horizon should be strongly suppressed.

The most recent discussion of anisotropies in the sky distribution of UHECRs began with the report that the arrival directions of the 27 cosmic rays observed by Auger with energies above 57 EeV exhibited a statistically significant correlation with the anisotropically distributed galaxies in the 12th VCV \cite{VC06} catalog of active galactic nuclei (AGN)  \cite{Auger1,Auger2}. The correlation was most significant for AGN with redshifts $z <$ 0.018 (distances $< $ 75 Mpc)  and within 3.1$^\circ$ separation angles. An independent dataset confirmed the anisotropy at a confidence level of over 99\% \cite{Auger1,Auger2}. The prescription established by the Auger collaboration tested the departure from isotropy given the VCV AGN coverage of the sky, not the hypothesis that the VCV AGN were the actual UHECR sources. A recent update of the anisotropy tests with 69 events above 55 EeV \cite{Abreu10} shows that the correlation with the VCV catalog is not as strong for the same parameters as the original period (20 events correlate out of the original 27 while only 12 correlate out of the new 42). The data after the prescription period shows a departure from isotropy at the 3$\sigma$ level. In this meeting it was shown that of the 84 Auger events above 55 EeV (after the 14 used for the prescription), 28 correlate \cite{AugerAnisICRC} which amounts to a (33 $\pm$ 5)\% correlation versus  21\% expected from isotropy. The Telescope Array showed that 8 out of 20 events correlate \cite{TA}, which is a 40\% correlation while  24\% is expected from isotropy. The two observations are consistent and show that an anisotropy signal is weak at these energies probably due to a large isotropic background. The lack of statistics at higher energies limits the reach of current observatories to achieve a clear detection if the anisotropy is due to the large scale structure and primaries are heavier than proton.

The anisotropy reported by the test with the VCV catalog  may indicate the effect of the large scale structure in the distribution of source harboring galaxies or it may be due to a nearby source. An interesting possibility is the cluster of Auger events around the direction of Centaurus A, the closest AGN (at $\sim$  3.8 Mpc). The most significant excess is in a 24 degree window around Cen A, where 19 of the 98 events are found, while  7.6 are expected by chance correlation  \cite{AugerAnisICRC}. The significance for the excess region can only be established with independent data. Only much higher statistics will tell if Cen A is the first UHECR source to be identified. 

\subsection{Composition}

The third key measurement that can help resolve the mystery behind the origin of UHECRs is their composition as a function of energy observed on Earth. Composition measurements can be made directly up to energies of $\sim$ 100 TeV with space-based experiments. For higher energies, composition is derived from the observed development and particle content of the extensive air shower created  by the primary cosmic ray when it interacts with the atmosphere. 

\begin{figure}[!t]
\centerline{\includegraphics[width=0.55\textwidth]{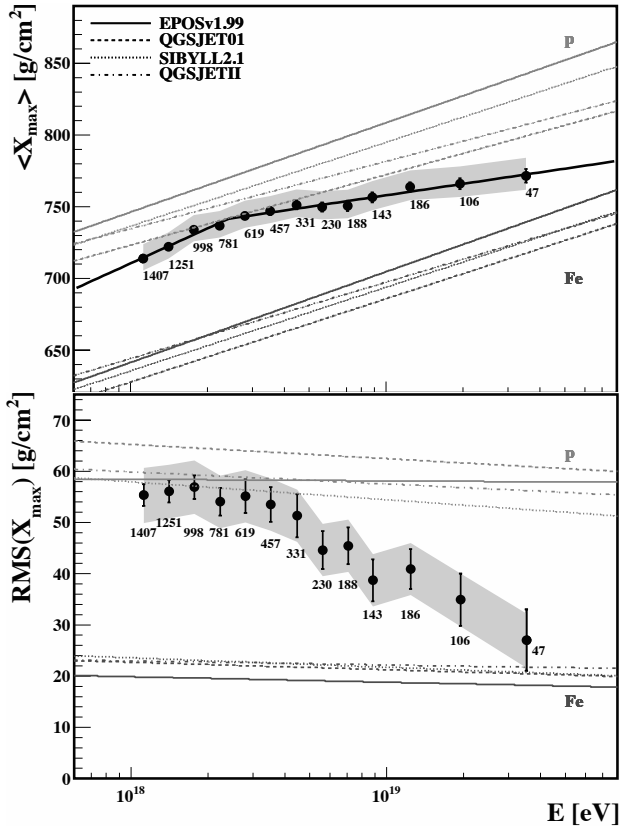}}
\caption{Average $X_{max}$ (top panel) and the RMS of $X_{max}$ (bottom panel)
are shown as a function of the energy. Auger data are the black points with statistical error bars. Systematic uncertainties are indicated as a grey band.
Predictions from different hadronic interaction models (EPOSv1.99 in solid lines, QGSJetII-03 in dash-dot lines,  and SIBYLL 2.1 in dotted lines) for proton ({p}) and iron (Fe) primaries are shown as labelled. }
\label{AugerComp}
\end{figure}

Assuming that hadronic interactions models describe reasonably well the air shower properties of different primaries at these energies, observations show the dominance of light nuclei around a few EeV. As shown in figure \ref{AugerComp}, a surprising trend occurs in data by the Auger Observatory above 10 EeV, a change toward heavy primaries is seen both in average position of the maximum of the showers  as well as in the rms fluctuations about the mean up to 40 EeV \cite{AugerCompICRC,Abraham:2010yv}. As a mixture of different nuclei would increase the rms fluctuations, the observed narrow distribution argues for a change toward a composition dominated by heavier nuclei. Using complementary techniques (asymmetry of the signal rise time and muon production depth) designed to make use of the high statistics of surface detector events, the Auger collaboration extended the measurement of shower properties to energies up to 60 EeV \cite{AugerCompICRC}. The trend toward heavier nuclei continues. The preliminary TA  measurement of fluctuations  remains closer to light primaries up to around 50 EeV. The two results are consistent within quoted errors, so the situation is currently unclear. 

As reported in this meeting, the study of showers of energies up to $10^{18.5}$ eV gives an estimate of the proton-air cross section of 505 mb $\pm 22$ (statistical) and $+$ 28 $-$ 36 (systematic uncertainties) \cite{AugerHadrICRC}.  Changes to hadronic interactions from current extrapolations provide a plausible alternative interpretation to the observed shower development behavior above $10^{18.5}$ eV. Auger probes interactions above 100 TeV center of mass, while hadronic interactions are only known around a TeV. The observation of anisotropies and secondary particles (neutrinos and gamma-rays) can lead to astrophysical constraints on the composition of UHECRs, opening the possibility for the study of hadronic interaction cross sections, multiplicities, and other interaction parameters at hundreds of TeV.

The detailed composition of UHECRs is still to be understood, but it is clear that primaries are not dominated by photons \cite{Aglietta:2007,Abraham:2009qb} or neutrinos \cite{Auger_nu09,Abbasi08neu}. Limits on the photon fraction place stringent limits on models where UHECRs are generated by the decay of super heavy dark matter and topological defects. Unfortunately, the uncertainties on the UHECR source composition, spectrum, and redshift evolution translates to many orders of magnitude uncertainty in the expected cosmogenic neutrino flux as discussed next.

 \section{The Galactic to extragalactic transition}\label{section:transition}

The highest energy cosmic rays are likely to originate in extragalactic sources, given the strength  of Galactic magnetic fields and the lack of correlations with the Galactic plane. Low energy cosmic rays are easily created and contained in the Galaxy, so a transition region should occur in some intermediate energy.  Modern measurements of the spectrum place a plausible transition region around the ankle at a few EeV (figures \ref{CRspec} and \ref{uhecrSpec}). However, the ankle can also be interpreted as the product of propagation losses due to pair production \cite{BG88,BGG06} in proton dominated scenarios allowing for a transition at lower energies.

The knee in the cosmic ray spectrum is likely to signal the maximum energy, $E_{\rm max}$, for light nuclei of dominant Galactic sources  and/or the maximum containment energy for light nuclei in the Galactic magnetic field. The same effect for heavier nuclei may cause  the softer spectrum above the knee (see, e.g., \cite{L05,Hillas06}). Extragalactic sources producing spectra harder than $s=3$  can overtake the decaying Galactic  flux around the ankle.  Recent studies of a transition at the ankle which fit the observed spectrum and the composition trends in this energy region are discussed in \cite{Allard05} where different models are contrasted. Models based on proton primaries with a hard spectrum \cite{WW04}, on a mixed composition with proportions similar to the Galactic mix, or even on a composition dominated by heavy nuclei \cite{Allard07} fit well the UHECR spectrum and composition data around the ankle. In figure \ref{uhecrSpec}, we show two examples of the so-called ``ankle transition models": one with source injection $s=2.1$, source composition similar to the Galactic mixture, and source evolution that follows the star formation rate (SFR); and a second model with similar source evolution and $s=2$, but a pure iron composition injected at the source. Both models fit well the UHECR spectrum but predict different compositions throughout this energy range.

Ankle transition models work well for UHECR scenarios, but they were thought to challenge models for the origin of Galactic cosmic rays. The requirement that Galactic sources reach energies close to the ankle strained traditional models where  acceleration  in supernova remnants (SNRs) was expected to  fade around 1 PeV  \cite{LC83}. A modification to the traditional SNR scenario, such as magnetic field amplification in SN shocks \cite{BL01}, or a different progenitors  such as Wolf-Rayet star winds \cite{Bierm93}, and trans-relativistic supernovae  \cite{Budnik08} may explain the energy gap from PeV to EeV.  Taking into account magnetic field amplification and Alfvenic drift in shocks of Type IIb SNRs, \cite{Ptuskin10} find that Galactic cosmic ray iron can reach $E_{\rm max}\sim$ 5 EeV, allowing extragalactic cosmic rays to begin to dominate above the ankle.

The possibility that the ankle is due to pair-production losses during the propagation of extragalactic protons \cite{BG88}  has motivated an alternative model for the Galactic to extragalactic transition, called ``dip models"  \cite{BGG06}. The energy of the predicted dip is close to the observed ankle and a good fit to the spectrum over a large energy range is reached with a softer injection index as the Òdip protonÓ models shown in figure \ref{uhecrSpec}. This option relaxes the need for Galactic cosmic rays to reach close to EeV energies, however it needs to be tuned to avoid strong spectral features between the knee and the ankle. Detailed models where the lower energy behavior of the extragalactic component blends smoothly with the Galactic cosmic rays have been developed using minimum energy and magnetic effects \cite{L05,AB05,Hillas06,KL08a,Globus08}.  
In some of these models a feature is produced around the ``second knee'' which may be observed around  0.5 EeV.  The dip model can fit the observed spectrum if the injection is proton dominated \cite{BGG05,Allard07} or with at most a primordial proton to helium mix \cite{Hillas06}, which gives a clear path for distinguishing it from mixed composition models. A proton dominated flux below the ankle region is a necessary condition for this model to be verified.

Clarifying the structure of the transition region is important for reaching a coherent picture of the origin of Galactic and extragalactic cosmic rays.  This requires accurate spectrum and composition measurements from the knee to the ankle and beyond. KASCADE-Grande \cite{KASCADE-Grande11} has made great progress above the knee, recently reporting an interesting structure in the composition. When dividing their sample into electron-rich and electron-poor, they find a kneelike structure in the heavy (electron-poor) component of cosmic rays around $8 \times 10^{16}$ eV\cite{KascGrandeKnee}. This feature in the heavy component spectrum  mimics the light (proton) knee structure at  $\sim 3 \times 10^{15}$ eV, giving credence to a rigidity dependent end of the Galactic cosmic rays.

UHECR projects have started to lower their energy threshold such as the Auger Observatory enhancements: HEAT (High Elevation Auger Telescopes)  \cite{AugerHEAT11} and AMIGA (Auger Muons and InÞll for the Ground Array)  \cite{AugerAMIGA11}; and the Telescope Array Low Energy Extension (TALE) \cite{TALE}. Having the same system covering a large range in energy will help control systematic offsets that degrade the accuracy of the needed precision. In addition, a strong multi-wavelength program  has shown that magnetic field amplification occurs in SNRs  and Galactic sources can reach further than previously believed. Finally, models of hadronic interactions will benefit from the energy reach of the LHC which can probe hadronic interactions at energies higher than the knee (figure \ref{CRspec}) and help constrain composition indicators between the knee and the ankle.

\section{Candidate Sources of UHECRs}\label{section:sources}

The requirements for astrophysical objects to be sources of UHECRs are quite stringent. Sources should be able to accelerate particles to above 100 EeV with high enough luminosities to explain the observed flux. The detailed shape of the observed flux and composition of UHECRs are not simply mapped onto what a candidate source injects, since the propagation from source to Earth modifies the spectrum, composition, and sky distribution of UHECRs. Propagation studies have become quite accurate when the effect of the relevant photon backgrounds is considered (including photons fro the cosmic microwave background up to ultraviolet background). However, magnetic fields are crucial for an accurate description of cosmic ray propagation and the magnitude and structure of cosmic magnetic fields is still quite uncertain.

\begin{figure}[!t]
\centerline{\includegraphics[angle=-90,width=0.6\textwidth]{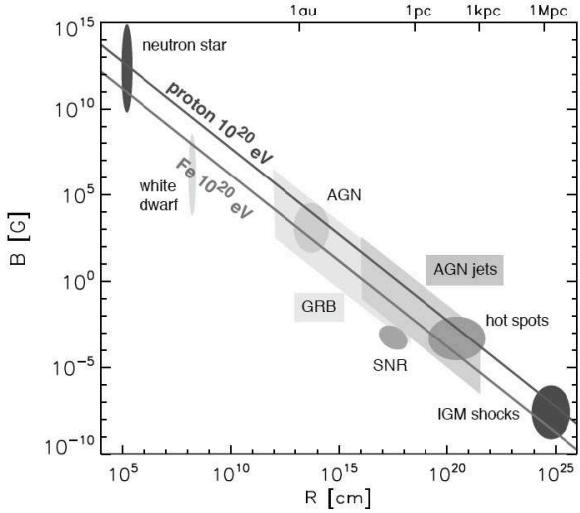}}
\caption{Hillas diagram of magnetic fields versus size of candidate UHECR sources. Above the diagonal lines protons or iron nuclei (as labelled) can be confined to a maximum energy of $E_{\rm max}= 10^{20}$~eV.  The most powerful candidate sources are shown with the uncertainties in their parameters. }
\label{HillasPlot}
\end{figure}

The Larmor radius of UHECRs in Galactic magnetic fields, $r_{\rm L} = E/ZeB \sim$ 110 kpc $Z^{-1} (\mu G/ B) (E/100$ EeV),  is much larger than the thickness of the Galactic disk. Thus, confinement in the Galaxy is not maintained at the highest energies, motivating  the search for extragalactic candidate sources.  Requiring that candidate sources be capable of confining particles up to $E_{\rm max}$, translates into a simple selection criterium for candidate sources with magnetic field strength $B$ and extension $R$  \cite{Hillas84}: 
$r_{\rm L} (E_{\rm max}) \le R$, i.e., $({R}/{110~{\rm kpc}}) ({B}/{1~\mu\mbox{G}}) \ge Z \, (E_{\rm max}/100 {\rm EeV})$. 
Figure \ref{HillasPlot} shows a ``Hillas diagram" where candidate sources are placed in a $B-R$ phase-space, including the range of these parameters for a given system. Most astrophysical objects do not even reach the  confinement line for iron at $10^{20}$~eV. Source candidates that pass the Hillas requirement include neutron stars, Active Galactic Nuclei (AGN), Gamma Ray Bursts (GRBs), and accretion shocks in the intergalactic medium. 

The Hillas criterion is a necessary condition, but not sufficient. In particular, most UHECR acceleration models rely on time dependent environments and relativistic outflows where the Lorentz factor $\Gamma\gg 1$. In the rest frame of the magnetized plasma, particles can only be accelerated over a transverse distance $R/\Gamma$, which tightens the Hillas requirement. In shock acceleration models the efficiency of the accelerator also makes the criterion stricter. In addition, the maximum accessible energy also depends on details of the acceleration process. The acceleration time needs to be smaller than the escape time of particles from the acceleration region, the lifetime of the source, and the energy loss time due to expansion and/or to interactions with the ambient medium. Thus, very few candidates survive a more careful study.

In addition to being able to accelerate up to $E_{\rm max} \gsim $ 200 EeV, candidate UHECR accelerators should have luminosities that can account for  the observed fluxes. A simple estimate of the required luminosity can be done 
assuming that all sources have the same injection spectral index $s$, the same steady luminosity in cosmic rays above $10^{19}$~eV, $L_{19}$, and that they are distributed homogeneously in the Universe with a number density $n_{\rm s}$.  To account for the observed flux of UHECRs at $10^{19}$~eV, the main quantity at play, $n_{\rm s}L_{19}$. The flux  at  $10^{19}$~eV gives $(E^3{\rm d}N/{\rm d}E)_{10^{19} {\rm eV}} \sim 10^{24}~{\rm eV^2\,m^{-2}\,s^{-1}\,sr^{-1}}(n_{\rm s}10^{5}\,{\rm Mpc^{3}})(L_{19}/10^{42}\,{\rm erg/s})$, for the case of $s=2.3$ and $E_{\rm max}=10^{20.5}$~eV. For reference, the number density of normal galaxies in the Universe today is of order $10^{-2}$~Mpc$^{-3}$, and it drops to $10^{-9}-10^{-8}$~Mpc$^{-3}$ for the most powerful active galaxies.
For transient sources, this scaling can be translated into: $(\dot{n}_{\rm s}/10^{-9}\,{\rm Mpc^{-3}\,yr^{-1}})$ $(E_{\rm tot,19}/3\times 10^{53}\,{\rm ergs})$, where $\dot{n}_{\rm s}$ is the birth rate of the source and $E_{\rm tot,19}$ the total injected energy in cosmic rays above $10^{19}$~eV. 

Below we briefly discuss the main astrophysical sites where UHECRs may originate.

{\bf Gravitational accretion shocks} are the largest systems that meet the Hillas criterium.
The accretion of dark matter and gas produce shocks around these large structures of the Universe (clusters of galaxies, filaments, walls), where diffusive shock acceleration can happen. For clusters of galaxies,  the linear extension of the magnetized shock can reach $\sim 1-10$~Mpc and the magnetic field can be $\sim 1~\mu$G (see, e.g., \cite{vanWeeren10}). However that the strength of the magnetic field  upstream of the shock can be much smaller, as it was produced out of the weakly magnetized void. The detection of very high energy gamma rays from these shocks would better constrain these parameters. A  time-dependent numerical calculation that includes  energy losses due to interactions of protons with radiative backgrounds   shows that the maximum energy achievable by protons does exceed a few times $10^{19}$ eV in these systems \cite{Vannoni09}.

{\bf Active Galactic Nuclei}  are composed of an accretion disk around a central super-massive black hole and are sometimes associated with jets terminating in lobes (or hot spots) which can be detected in radio. For a black hole of mass $\sim 10^9~{\rm M}_\odot$, the equipartition magnetic field in the central region yields $B\sim 300$~G. Assuming the central region to be of order $R\sim 100$~A.U., particles could be confined up to $E_{\rm max} \sim $ 150 EeV and accelerated by electrostatic acceleration in the black hole magnetosphere (e.g., \cite{Boldt99}). This energy is hardly reached by particles in practice due to energy losses in this dense region. Radio loud galaxies could also accelerate particles in their inner jets (see e.g., \cite{Rieger07}). The quantity $B\ R\sim 0.3$~G~pc for the jets of a $\sim 10^9~{\rm M}_\odot$ black hole, leading to $E_{\rm max} \sim$ 300 EeV, but the acceleration is limited by photo-interactions and adiabatic losses making the escape of UHE particles non trivial \cite{Mannheim93}. 
For hot spots, the escape should be easier than in the jet (see, e.g., \cite{RB93}), but the acceleration in the bow shock is non trivial \cite{Berezhko08}. In addition, only the brightest (and rarest) AGN can meet the energetic requirements to steadily accelerate particles to the highest energies \cite{LW09}. Since the highest energy events do not point to these rare sources, the acceleration is likely to be transient phenomenon \cite{Farrar09} instead of continuos.

If UHECRs are accelerated in AGN,  the gamma-ray spectrum of these sources should display signatures of hadronic processes. Future gamma-ray telescopes (such as the Cherenkov Telescope Array) may distinguish these hadronic signatures from leptonic acceleration. 

{\bf Gamma-ray Bursts} may also be sources of UHECRs \cite{W95,V95}. 
The explosion of a GRB leads to the formation of multiple shock regions which are potential acceleration zones for UHECRs. The magnetic field at these shocks is $\sim 10^6~$G at a distance $R\sim 10^{12}$~cm from the center. These values are derived for internal shocks that happen before the ejected plasma reaches the interstellar medium, assuming $B\sim 10^{12}$~G near the central engine (of size $R\sim 10~$km) and an evolution $B\propto R^{-1}$. The wide region presented in figure \ref{HillasPlot} stems from the time dependence of the event. Models based on GRBs allow acceleration up to $\sim 10^{20}$~eV for selected choices of magnetic field strength and structure in the different shocks of the event. The flux of gamma-rays reaching the Earth from GRBs is generally comparable to the observed flux of UHECRs, implying a tight energetic requirement  for GRBs to be the sources of UHECRs. With a GRB rate of $\sim 0.3$~Gpc$^{-3}$~yr$^{-1}$ at $z=0$, it can be calculated that the energy injected isotropically (regardless of beaming) in UHECRs needs to be of order $E_{\rm UHECR}>10^{53}$~erg \cite{Guetta07,Zitouni08,Budnik08}. The transient nature of these objects can help explain the lack of powerful counterparts correlating with the arrival direction of the highest energy cosmic rays, however, the trend toward a heavy composition at the highest energies is a challenge for GRB models.

{\bf Neutron Stars} can easily fulfill the Hillas criterion and might prove to be very good candidate sources, though they are scarcely discussed in the UHECR literature. Magnetized rotating neutron stars  (i.e., pulsars) have been suggested as possible accelerators of cosmic rays since their discovery \cite{Gunn69}, due to their important rotational and magnetic energy reservoirs. Galactic pulsars have been suggested as the sources of cosmic rays around the knee region up to the ankle (see, e.g., \cite{BB04}).  Iron nuclei accelerated in the fastest spinning young neutron stars were used to explain the observed cosmic rays above the ankle in a UHECR Galactic source scenario \cite{Blasi00}.  The stripping of heavy nuclei from the surface of the star can seed the magnetized wind that accelerates UHECRs through a unipolar inductor. The final spectrum is a hard, $J \propto E^{-1}$, due to the spin down rate of young pulsars \cite{Blasi00}. The birth of extragalactic magnetars (neutron stars with  extremely strong surface dipole fields of $\sim 10^{15}$ G) was also proposed as a source of ultrahigh energy protons \cite{Arons03},  assuming that the magnetar birth generates a jet that breaks through the supernova envelope.

The proposals for the origin of UHECRs in young neutron stars of \cite{Blasi00} and \cite{Arons03} were elaborated to explain the absence of the GZK effect  in the observed spectrum reported by AGASA \cite{Takeda98} without invoking the so-called top-down models.  An increase in the exposure at the ultrahigh energies has shown that the UHECR spectrum is consistent with a GZK effect, but the composition may be heavier at the highest energies. This brings new interest in exploring neutron stars as candidate sources due to the ease of injecting large proportions of heavy nuclei into an acceleration region. The birth of pulsars is also a transient event, which makes a direct correlation between the highest energy events and the source unattainable, as discussed below.

\section{When should Cosmic Rays start to point?}\label{subsection:astronomy}
One of the most puzzling facts concerning UHECRs is the absence of clear sources in the arrival directions of the highest energy events. If  sources are powerful astrophysical accelerators, photon counterparts should be visible in the arrival direction at the highest energie cosmic rays. Current upper limits on the strength of cosmic magnetic fields suggest  that prontons should not be deflected by more than a few degrees above 60 EeV, thus some correlation should exist with the underlying baryonic matter, unless they are heavier nuclei (which postpones the onset of correlations by $Z$).

As a result, many authors have searched for correlations between existing data and astrophysical object catalogs. A few correlations have been reported over the years without a clear confirmation (see, e.g., \cite{TT01,Stanev95}).
The latest correlation result concerns the highest energy events ($E>$ 55 EeV) detected by the Auger Observatory and AGN within distance $<75$~Mpc \cite{Auger1,Auger2, Abreu10}. These results show that above 55 EeV the distribution of events are anisotropic with 99\% CL. The mild anisotropy (33 $\pm$ 5\%) is likely to be due to the large scale structures along which AGN are distributed. Another possible  interpretation is that Auger may be observing in part the last scattering surface of UHECRs rather than their source population  \cite{KL08b}. 

Another explanation for the absence of counterparts in the arrival direction of UHECRs could reside in the very nature of the sources. The delay induced by extragalactic magnetic fields of mean strength $B_{\rm nG}= B/ 10^{-9}\,{\rm G} $ and coherence length $\lambda_{\rm Mpc} = \lambda/{\rm Mpc} $ on particles of charge $Z$ and energy $E_{20} = E / 10^{20}$ eV with respect to photons over a distance $D_{\rm Mpc} = D /{\rm Mpc}$ reads \cite{AH78}:
\begin{equation}\label{eq:delay}
\delta t\,\simeq\,2.3\times 10^2\,{\rm yrs}\,
Z^2\left(\frac{D_{\rm Mpc}}{10}\right)^2\,\left(\frac{\lambda_{\rm Mpc}}{0.1}\right)
\,E_{20}^{-2} \, {B_{\rm nG}}^2 .
\end{equation}
For intergalactic magnetic fields of lower strength ($B \lsim10^{-12}$~G), the time delay can be shorter than a year over 100~Mpc. However, the crossing of one single magnetized filament (with thickness $r_f$ and field $B_f$) will lead to a deflection that induces a time delay with respect to a straight line of order: $\delta t_f\,\simeq\, 10^3\,{\rm yr}\,({r_f / 2\,{\rm Mpc}})^2 ({B_f/ 10\,{\rm nG}})^2\,({\lambda_{\rm Mpc}/0.1}) E_{20}^{-2}$.

For transient sources like GRBs, young neutron stars, or AGN flares which have an activity timescale $\ll \delta t$, this delay is sufficient to erase any temporal coincidence between UHECRs and their progenitors \cite{W95,V95}. 

As discussed above, UHECR sky anisotropies and their composition are tightly connected. In particular, if an anisotropy signal is measured above an energy $E_{\rm thr}$ assuming that it is produced by heavy nuclei of charge $Z$, one expects an anisotropy signal to be also present at energy $E_{\rm thr}/Z$ due to the proton component, depending on the proton to heavy nuclei ratio injected at the source or produced via propagation  \cite{LW09}. 

\section{Secondaries from UHECRs}

\begin{figure}[!t]
\centerline{\includegraphics[width=0.5\textwidth]{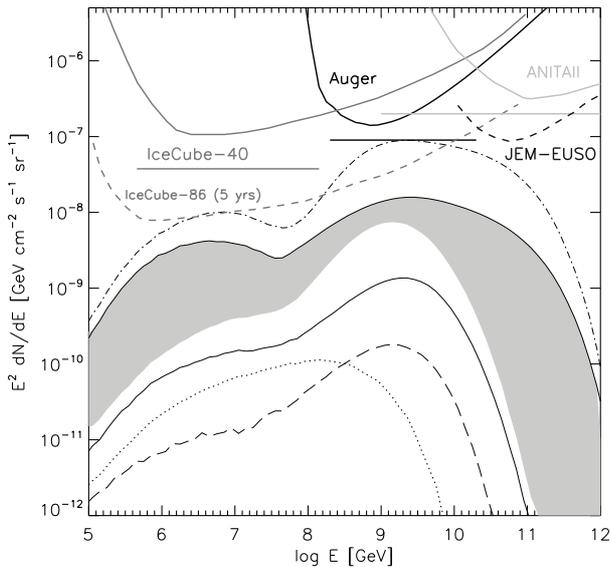}}
\caption{Cosmogenic neutrino flux for all flavors, for different UHECR parameters compared to instrument sensitivities (adapted from \cite{KAO10}). Dash-dotted line corresponds to a strong source evolution case (FRII evolution, see \cite{Wall05}) with a pure proton composition, dip transition model, and $E_{\rm max}=$ 3~ZeV. Uniform source evolution with: iron rich (30\%) composition and $E_{Z,\rm max}<Z$ 10 EeV is shown in the dotted line and the dashed line is for pure iron injection and $E_{Z,\rm max}=Z$ 100 EeV. Grey shaded range brackets dip and ankle transition models, with evolution of star formation history for $z<4$, pure proton and mixed `Galactic' compositions, and large proton $E_{\rm max} ( >  100$ EeV)). Including the uniform source evolution would broaden the shaded area down to the black solid line. Experimental limits (solid lines) assume 90\% confidence level and full mixing neutrino oscillation. The differential limit and the integral flux limit on a pure $E^{-2}$ spectrum (straight line) are presented for IceCube-22 \cite{Abbasi10}, Ice-Cube-40 \cite{Abbasi11}, ANITA-II \cite{ANITA10} and Auger \cite{Auger_nu09}. Dashed lines show future sensitivities for IceCube 80 lines \cite{Karle10}, and for JEM-EUSO  \cite{JemEUSO}).}
\label{cosmoNeut}
\end{figure}

Secondary neutrinos and photons can be produced by UHECRs when they interact with ambient baryonic matter and radiation fields inside the source or during their propagation from source to Earth. These particles travel in geodesics unaffected by magnetic fields and bear valuable information of the birthplace of their progenitors. The quest for sources of UHECRs has thus long been associated with the detection of neutrinos and gamma rays that might pinpoint the position of the accelerators in the sky.

The detection of these secondary particles is not straightforward however: first, the propagation of gamma rays with energy exceeding several TeV is affected by their interaction with CMB and radio photons. These interactions lead to the production of high energy electron and positron pairs which in turn up-scatter CMB or radio photons by inverse Compton processes, initiating electromagnetic cascades. As a consequence, one does not expect to observe gamma rays of energy above $\sim 100$~TeV from sources located beyond a horizon of a few Mpc. Above EeV energies, photons can again propagate over large distances, depending on the radio background, and can reach observable levels around tens of EeV. Secondary neutrinos are very useful because, unlike cosmic-rays and photons, they are not absorbed by the cosmic backgrounds while propagating throughout the Universe. In particular, they give a unique access to observing sources at PeV energies. However, their small interaction cross-section makes it difficult to detect them on the Earth requiring the construction of km$^3$ or larger detectors.

Neutrinos generated during UHECR propagation \cite{BZ69,Stecker79}, often called cosmogenic neutrinos, represent a ``guaranteed flux'' and have encouraged efforts to detect them for decades (see, e.g.,  \cite{AM09}). One important assumption for the existence of cosmogenic neutrinos, that cosmic rays are extragalactic at the highest energies, has been verified by the detection of a feature consistent with the GZK cutoff  in the cosmic ray spectrum \cite{Abbasi09,Abraham:2008ru}  and by the indication of anisotropies in the cosmic ray sky distribution at the highest energies  \cite{Auger1,Auger2}. 
These findings herald a possible resolution to the mystery behind the origin of UHECRs and the possibility of detecting ultrahigh energy neutrinos in the near future. 

This optimistic view has been dampened by the indication that UHECRs may be dominated by heavier nuclei \cite{AugerCompICRC,Abraham:2010yv}. The cosmogenic neutrino flux expected from heavy cosmic ray primaries can be much lower than if the primaries are protons at ultrahigh energies, making a detection extremely challenging for current observatories.  Conversely, if neutrinos are observed, they will test specific sets of cosmic ray source parameters.

Figure~\ref{cosmoNeut} summarizes the effects of different assumptions about the UHECR source evolution, the Galactic to extragalactic transition, the injected chemical composition, and $E_{\rm max}$, on the cosmogenic neutrino flux (adapted from \cite{KAO10}). It demonstrates that the parameter space is  poorly constrained with uncertainties of several orders of magnitude in the predicted flux.

Due to the delay induced by cosmic magnetic fields on charged cosmic rays, secondary neutrinos and photons should not be detected in time coincidence with UHECRs if the sources are not continuously emitting particles, but are transient such as gamma-ray bursts and young pulsars.

\section{Discussion}

The resolution of the long standing mystery of the origin of ultrahigh energy cosmic rays will require a coordinated approach on three complementary fronts: the direct ultrahigh energy cosmic ray frontier, the transition region between the knee and the ankle, and the multi-messenger interface with high-energy photons and neutrinos.

\begin{figure}[!t]
\centerline{\includegraphics[angle=270,width=0.5\textwidth]{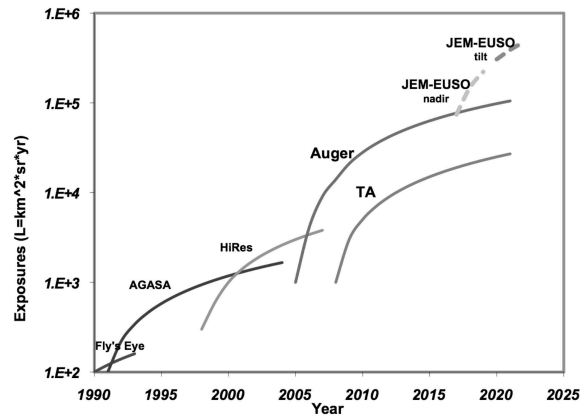}}
\caption{Exposures to UHECRs from 1990 to present from Fly's Eye, AGASA, HiRes, Auger and TA. Estimates of the exposures in the future of Auger and TA plus the planned space observatory,  JEM-EUSO \cite{JemEUSO}).}
\label{ExposFutur}
\end{figure}

Current data suggest that watershed anisotropies will only become clear above 60 EeV and that very large statistics with good angular and energy resolution will be required. The Auger Observatory (located in Mendoza, Argentina),  will add $7\times10^3 $ km$^2$ sr each year of exposure to the southern sky, while the Telescope Array (located in Utah, USA) will add about $2\times10^3 $ km$^2$ each year in the North as shown in figure \ref{ExposFutur}. Current technologies can reach a goal of another order of magnitude if deployed by bold scientists over very large areas. New technologies may ease the need for large number of detector units to cover similarly large areas. 

A promising avenue to reach the necessary high statistics is the idea of space observatories (e.g., JEM-EUSO, OWL, Super-EUSO). With current technologies, a large statistics measurement of the spectrum and angular distribution of arrival directions above GZK energies are well within reach. 
Improved photon detection technologies will be needed to reconstruct shower maxima from space. If deployed in 2017, JEM-EUSO can significantly increase the exposure to UHECRs reaching the level needed to unveil this mystery \cite{JemEUSO} as  in figure \ref{ExposFutur}.

With a coordinated effort, the next generation observatories can explore more of the $\sim 5$~million trans-GZK events the Earth's atmosphere receives per year and find the highest energy accelerators in the Universe.

\section{Acknowledgements}
Thanks to K. Kotera and the Auger group  for very fruitful discussions. This work was supported by the NSF grant PHY-1068696 at the University of Chicago, and the Kavli Institute for Cosmological Physics through grant NSF PHY-1125897 and an endowment from the Kavli Foundation.

\clearpage


\begin{thebibliography}{}

\bibitem{Aharonian}
F. Aharonian, Proc. 32nd ICRC, 2011, in these Proceedings

\bibitem{KKAO11}
K.~Kotera, A.~V. Olinto, Annu. Rev. Astron. Astrophys. 2011, 49, 119

\bibitem{Letessier11}
A.~{Letessier-Selvon}, T.~{Stanev}, Rev. Mod. Phys., 2011, 83, 907

\bibitem{G66}
K.~Greisen, Phys. Rev. Lett., 1966, 16, 748

\bibitem{ZK66}
G.~Zatsepin, V.~Kuzmin,  J. Exp.  Theor. Phys. Lett., 1966, 4, 78

\bibitem{Ahn08}
H.~S.  {Ahn}, et al., (ATIC Collab) International Cosmic Ray Conference, 2008, 2,  79

\bibitem{Grigorov71}
N.~L. {Grigorov}, et al. (PROTON Collab), International Cosmic Ray Conference, 1971, 1, 172
 
\bibitem{Apanasenko01}
A.~V {Apanasenko}, et al. (RUNJOB Collab), International Cosmic Ray Conference, 2001, 5, 1622 

\bibitem{Chen08}
D. Cheng et al. (Tibet AS-$\gamma$), International Cosmic Ray Conference, 2008, 4,  103 

\bibitem{Kampert04}
K.-H. {Kampert},  et al. (KASCADE Collab), Nuclear Physics B Proceedings Supplements, 2004,  136, 273

\bibitem{KASCADE-Grande11} 
W.~D. {Apel}, et al., Proc. 32nd ICRC, 2011, 1307  (these proceedings)

\bibitem{Abbasi09}
R.~U. {Abbasi}, et~al., Astroparticle Physics, 2009, 32, 53. 

\bibitem{Abbasi08}
R.~U. {Abbasi}, et al. (HiRes Collab), Phys. Rev. Lett., 2008, 100, 10

\bibitem{Abraham10}
J.~{Abraham}, et~al., Phys. Lett. B, 2010, 685, 239

\bibitem{Abraham:2010yv}
J.~{Abraham}, et~al., Phys. Rev. Lett., 2010, 104,  091101

\bibitem{Linsley63}
J. Linsley, Phys. Rev. Lett., 1963, 10, 146

\bibitem{Chiba92}
N. Chiba et al., Astropart. Phys., 1992, 1, 27

\bibitem{Boyer02}
J. H. Boyer, Nucl. Instr. and Meth. in Phys. Res., 2002, A482, 457-474

\bibitem{Bird:1994}
D. J. Bird et al., Astrophys. J., 1994, 424, 491-502

\bibitem{Baltrusaitis85}
R. Baltrusaitis et al., Nucl. Instr. and Meth. in Phys. Res., 1985, A240, 410
 
\bibitem{Abraham04}
J. Abraham et al., Nucl. Instrum. Meth. in Phys. Res. A, 2004, 523, 50-95
 
\bibitem{TA}
Y. Tsunesada et al.  (Telescope Array Collaboration),  Proc. 32nd ICRC, 2011, in these Proceedings

\bibitem{AugerSpecICRC}
F. Salamida for the Pierre Auger Collaboration, Proc. 32nd ICRC, 2011, in these Proceedings

\bibitem{Abraham:2008ru}
J.~{Abraham}, et~al., Phys. Rev. Lett., 2008, 101,  061101.


\bibitem{Takeda98}
M.~{Takeda}, et~al., Physical Review Letters, 1998, 81, 1163


\bibitem{Allard07}
D.~{Allard}, E.~{Parizot}, A.~V. {Olinto}, Astroparticle Phys., 2007, 27,  61

\bibitem{BG88}
V.~S. {Berezinsky}, S.~I. {Grigorieva},  A\&A, 1988, 199, 1

\bibitem{BGG06}
V.~{Berezinsky}, A.~{Gazizov}, S.~{Grigorieva}, Phys. Rev. D,  2006, 74~(4), 043005


\bibitem{VC06}
M.-P. {V{\'e}ron-Cetty}, P.~{V{\'e}ron},  A\&A, 2006, 455, 773

\bibitem{Auger1}
J.~{Abraham}, et~al.,  Science, 2007,  318  938

\bibitem{Auger2}
J.~{Abraham}, et~al.,  Astroparticle Physics, 2008, 29,  188

\bibitem{Abreu10}
P.~{Abreu}, et~al., Astroparticle Physics,  2010, 34, 314


\bibitem{AugerAnisICRC} 
K.-H. Kampert for the Pierre Auger Collaboration, Proc. 32nd ICRC, 2011, in these Proceedings (Volume of invited and highlight papers)


\bibitem{AugerCompICRC} 
D. Garcia Pinto for the Pierre Auger Collaboration, Proc. 32nd ICRC, 2011, in these Proceedings, contribution 709


\bibitem{AugerHadrICRC}
R. Ulrich for the Pierre Auger Collaboration, Proc. 32nd ICRC, 2011, in these Proceedings, contribution 946


\bibitem{Aglietta:2007}
J.~Abraham, et~al.,  Astropart.  Phys., 2008, 29, 243

\bibitem{Abraham:2009qb}
J.~{Abraham}, et~al., Astropart. Phys., 2009,  31,  399

\bibitem{Auger_nu09}
J.~{Abraham}, et~al., Phys.~Rev., 2009, D 79, 102001

\bibitem{Abbasi08neu}
R.~U. {Abbasi}, et~al., Astrophys. J., 2008, 684,  790

\bibitem{L05}
M. Lemoine, Phys. Rev.,  2005,  D 71, 8

\bibitem{Hillas06}
A.~M. Hillas, 2006,  arXiv:astro-ph/0607109

\bibitem{Allard05} 
D. Allard, E. Parizot, A.~V. {Olinto}, E. {Khan},  and S. {Goriely},  A\&A, 2005, 443,  L29-L32

\bibitem{WW04}
T. {Wibig}  and A.~W.{Wolfendale}, 2004, arXiv:astro-ph/0410624
 
\bibitem{Allard07}
D. {Allard}, E. {Parizot}, and A.~V.{Olinto}, Astroparticle Physics, 2007, 27, 61-75

\bibitem{LC83}
 P.~O. {Lagage},and C.~J. {Cesarsky}, A\&A, 1983, 125, 249-257

\bibitem{BL01}
A.~R. Bell,  and  S.~G.{Lucek}, MNRAS, 2001, 321, 433-438

\bibitem{Bierm93}
P.~L. {Biermann},  and J.~P. {Cassinelli}, Astron.~Astrophys., 1993, 277, 691

\bibitem{Budnik08}
R.  {Budnik}, B. {Katz}, A. {MacFadyen},  and E. {Waxman}, ApJ, 2008, 673, 928-933

\bibitem{Ptuskin10}
 V.  {Ptuskin}, V.  {Zirakashvili},  and  E.-S. {Seo},  Ap. J., 2010, 718, 31-36


\bibitem{AB05}
R. Aloisio and V.~S.{Berezinsky}, ApJ, 2005, 625, 249-255

\bibitem{AH78}
C. Alcock and S. Hatchett,  Astrophys.~J., 1978, 222, 456-470

\bibitem{KL08a}
K. {Kotera}  and M. {Lemoine}, Phys. Rev. D, 2008, 77  - 2, 023005

\bibitem{Globus08} 
N. {Globus},  D. {Allard}, and E. {Parizot}, A\&A, 2008, 479, 97-110

\bibitem{BGG05}
 V. {Berezinsky},  A.{Gazizov}, and  S. {Grigorieva}, Phys.Lett.B, 2005, 612, 147

\bibitem{KascGrandeKnee}
W.~D. {Apel}, et al., Phys. Rev. Lett., 2011, 107, 17 

\bibitem{AugerHEAT11}
H. J. Mathes for the Pierre Auger Collaboration, Proc. 32nd ICRC, 2011, 1307, 0761 (in these Proceedings)  

\bibitem{AugerAMIGA11}
F. Sanchez for the Pierre Auger Collaboration, Proc. 32nd ICRC, 2011, 1307 , 0742 (in these Proceedings)

\bibitem{TALE}
T. Shibata et al., Proc. 32nd ICRC, 2011, 1307  (in these Proceedings)

\bibitem{Hillas84}
A.~M. {Hillas},  ARAA,  1984, 22, 425-444

\bibitem{vanWeeren10} 
R.~J. van Weeren, H.~J.~A. {Ršttgering},  M.  {BrŸggen}, and M. {Hoeft}, Science,  2010, 330, 347

\bibitem{Vannoni09}
G. {Vannoni},  F.~A. {Aharonian}, S. {Gabici},  S.~R. {Kelner}, and A. {Prosekin}, Astronomy \& Astrophysics, 2011, 536, A56

\bibitem{Boldt99}
E. Boldt and P. {Ghosh}, MNRAS, 1999, 307, 491-494


\bibitem{Rieger07}
F.~M.{Rieger},  V.{Bosch-Ramon},  and P. {Duffy},  Ap \& SS, 2007, 309, 119-125

\bibitem{Mannheim93}
K. Mannheim, A\&A, 1993, 269, 67-76

\bibitem{RB93}
J.~P. Rachen  and P.~L. {Biermann}, A\&A, 1993, 272, 161


\bibitem{Berezhko08}
E.~G. {Berezhko}, ApJ Letters, 2008, 684, L69-L71
 
\bibitem{LW09}
M. Lemoine and E. Waxman, JCAP, 2009, 0911, 009
 
\bibitem{Farrar09} 
G.~R. {Farrar}  and A. {Gruzinov}, ApJ, 2009, 693, 329-332

\bibitem{W95}
E. {Waxman}, Physical Review Letters,  1995, 75, 386-389

\bibitem{V95}
M. {Vietri}, ApJ, 1995, 453, 883

\bibitem{Guetta07}
D. Guetta and T. Piran,  JCAP, 2007, 7, 3

\bibitem{Zitouni08}
H. {Zitouni},  F. {Daigne}, R. {Mochkovich},  and T.~H. {Zerguini},  MNRAS, 2008, 386, 1597-1604


\bibitem{Gunn69}
J.~E.{Gunn} and J.~P. {Ostriker} , Physical Review Letters, 1969, 22, 728-731

\bibitem{BB04}
W. Bednarek and M. Bartosik, A\&A, 2004, 423, 405-413

\bibitem{Blasi00} 
P. {Blasi}, R.  {Epstein}, and A.V. {Olinto}, Astrophys. J. Lett., 2000, 533, L123-L126

\bibitem{Arons03} 
J. Arons, Astrophys. J., 2003, 589, 871-892

\bibitem{TT01}
P.~G. {Tinyakov} and I.~I.{Tkachev}, JETP Lett., 2001, 74, 1-5

\bibitem{Stanev95}
T. Stanev,  Phys. Rev. Lett., 1995, 75, 3056-3059

\bibitem{KL08b}
K. Kotera and M. Lemoine, Phys. Rev. D, 2008, 77, 123003
 
\bibitem{JemEUSO}
Y.~{Takahashi}, et~al., {The Jem-Euso Mission}, New J. Phys., 2009,  11, 065009

\bibitem{BZ69}
V.~S. {Berezinsky}, G.~T. {Zatsepin}, Physics Letters, 1969, B 28, 423

\bibitem{Stecker79}
F.~W. {Stecker},  Astrophys. J., 1979, 228  919

\bibitem{KAO10}
K.~{Kotera}, D.~{Allard}, A.~V. {Olinto}, JCAP, 2010, 10, 13

\bibitem{Wall05}
J.~V. {Wall}, C.~A. {Jackson}, P.~A. {Shaver}, I.~M. {Hook}, K.~I.
 {Kellermann},  A\&A, 2005, 434, 133

\bibitem{Abbasi10}
R.~U. {Abbasi}, et~al., Physical Review Letters, 2010,  104,  161101

\bibitem{Abbasi11}
R.~U. {Abbasi}, et~al., Physical Review, 2011, D 84, 082001 

\bibitem{ANITA10}
P.~W. {Gorham}, et~al.,  arXiv:1003.2961

\bibitem{Karle10}
A.~{Karle}, {IceCube}, arXiv: 1003.5715

\bibitem{AM09}
L.~A. {Anchordoqui}, T.~{Montaruli}, Ann. Rev. Nucl. Part. Sci., 2009, 60, 129

\end{thebibliography}
\end{document}